\documentclass[
 reprint,
 amsmath,amssymb,
]{revtex4-2}

\usepackage{graphicx}
\usepackage{dcolumn}
\usepackage{bm}
\usepackage{braket}
\usepackage{color}
\usepackage{txfonts}

\begin{document}

\title{Dynamics of Interacting Bosons on the Sawtooth Lattice with a Flat Band}

\author{Ko Gondaira$^1$, Nobuo Furukawa$^1$, and Daisuke Yamamoto$^2$}
\affiliation{$^1$Department of Physics and Mathematics, Aoyama Gakuin University, Sagamihara, Kanagawa 252-5258, Japan}
\affiliation{$^2$Department of Physics, College of Humanities and Sciences,
Nihon University, Sakurajosui, Setagaya, Tokyo 156-8550, Japan}

\begin{abstract}
    {Quantum many-body systems are expected to relax to a thermal state over time, with some exceptions such as systems with atypical eigenstates.  In this study, we investigate the effect of the existence of spatially localized eigenstates on the relaxation dynamics of interacting bosons loaded into a one-dimensional sawtooth lattice, which exhibits a flat band in the single-particle spectrum by tuning the hopping rates. Using the time-evolving block decimation algorithm, we simulate the time evolution of the local density profile based on the Bose-Hubbard model with different initial conditions. Our results show the presence of the flat band leads to a significant slowing down of relaxation for weak interactions. Even for strong interaction, when the initial state includes an isolated localized single-particle eigenstate in the superposition, remnants of the initial bias in the density profile persist for a long time. This particular relaxation dynamics can be tested using ultracold atoms in optical lattices.}
\end{abstract}


\maketitle

\section{\label{sec:intro}Introduction}
Unitary time evolution of an isolated quantum system is governed by the Schr\"odinger equation. However, predicting the future evolution of general quantum many-body systems is difficult due to the vast dimensionality of the state space. Nevertheless, macroscopic objects composed of a huge number of particles exhibit universal features in time evolution, such as thermal equilibration. The eigenstate thermalization hypothesis (ETH) is a well-known hypothesis that attempts to explain the mechanisms of thermal equilibration from a microscopic perspective. According to the ETH, each of energy eigenstates of an isolated many-body system produces the same expectation values of local observables as those in the microcanonical ensemble at the same energy \cite{deutsch2018eigenstate,abanin2019colloquium}.  Numerical evidence of the ETH has been found in various non-integrable lattice models, although some exceptions exists, including many-body localization (MBL) \cite{nandkishore2015many} and quantum many-body scars \cite{turner2018weak}. MBL is an emergent many-body phenomena where most of the eigenstates violate the ETH due to, e.g., strong disorder, and local information in the initial state can persist for a long time during time evolution \cite{nandkishore2015many}. Quantum many body scars are atypical eigenstates that comprise only a fraction of all eigenstates and avoid thermal equilibration for certain initial states \cite{turner2018weak}.

Although those non-equilibrium phenomena have been primarily discussed through numerical calculations on lattice models, recent developments in the control of artificial quantum systems, such as ultracold atoms \cite{bloch2008many}, trapped ions \cite{blatt2012quantum}, and nuclear and electron spins associated with impurity atoms in diamond \cite{doherty2013nitrogen, schirhagl2014nitrogen}, have enabled experimental realization and verification in the laboratory \cite{abanin2019colloquium, turner2018weak}. For example, in a system of ultracold atoms simulating a one-dimensional Fermi-Hubbard chain with quasi-random disorder, the MBL was observed in the relaxation dynamics of the initial charge density wave state \cite{schreiber2015observation}. In an experiment with Rydberg atoms loaded into an optical tweezer array, periodic revivals were observed for certain initial states while the other generic states showed relaxation to thermal equilibrium \cite{serbyn2021quantum, bernien2017probing}. This behavior has been explained by quantum many-body scars \cite{turner2018weak}.

The geometry of the lattice can be a significant factor affecting the dynamics of quantum systems~\cite{danieli2020many,mcclarty2020disorder}. By tuning the strength of particle hoppings on specific lattice structures \cite{Huber2010-bw}, one can create flat bands in the dispersion relation of the single-particle excitation spectrum. The existence of the flat band allows for the construction of spatially localized eigenstates that can inhibit the relaxation of certain initial states. In Refs. \cite{kuno2020flat,kuno2020flatscar,kuno2021multiple}, disorder-free MBL-like behavior and the existence of scars have been numerically discussed through simulations of the relaxation dynamics in flat-band systems of low density spinless fermions with nearest-neighbor interactions. In Ref. \cite{khare2020localized}, the quench dynamics and its localization behavior of hardcore bosonic flat-band systems with nearest-neighbor interactions have been studied.

In this paper, toward a deeper understanding and experimental realization of the unusual dynamics of flat-band systems, we examine the relaxation dynamics of a standard Bose-Hubbard system of interacting bosons with nearest-neighbor hopping and onsite interactions on sawtooth lattice \cite{zhang2015one} for different initial conditions. {The sawtooth lattice is formed by a one-dimensional series of corner-shared triangles with two types of hopping, and it is known that fine-tuning their ratio yields a flat band in the single-particle spectrum, associated with the existence of localized eigenstates. As for the initial conditions, we consider two scenarios: (i) the quarter-filled state where the particles are distributed at even intervals and (ii) the half-filled state where the particles are laid out in the left half of the system.} 

One reason for exploring such a model is its potential for experimental realization in cold-atom systems. {The standard Bose-Hubbard model includes only nearest-neighbor hopping and onsite interactions, which does not require long-range interacting particles, such as dipolar atoms, for realization in cold-atom experiments. Moreover, the initial states we study here could be prepared in experiments. For case (i), it is potentially achievable to prepare the initial state by employing techniques designed to create a density wave state with the assistance of an additional optical lattice~\cite{Trotzky2012}. The initial state in case (ii) could be prepared by optically removing the right half of the atomic population~\cite{choi2016exploring}.} 
{Investigating the softcore bosonic flat-band system is also of fundamental scientific interest.} Unlike in the fermionic or hardcore cases \cite{kuno2020flat,kuno2020flatscar,kuno2021multiple,khare2020localized}, a large number of localized single-particle eigenstates can still be found in the non-interacting limit even at high density due to the absence of the Pauli blocking. Therefore, it is nontrivial how the dynamical behavior of the bosonic flat-band system changes due to the softcore nature under the presence of interactions.

We investigate the dynamics of the system utilizing the time-evolving block decimation (TEBD) method~\cite{paeckel2019time,fishman2022}. This approach handles the time evolution of the matrix product state (MPS) through the Trotter-Suzuki decomposition. We first examine the single-particle and few-particle dynamics to get some analytical insights, followed by the numerical considerations on many-body dynamics. 

 We will demonstrate {for case (i)} that the spatial inhomogeneity of the initial state does not completely disappear, even after a long time and even with strong interactions. This is because the exact localized eigenstates are not eliminated from the superposition components of the state and survive in time evolution. {Case (ii)} is of particular interest since the speed of the relaxation strongly depends on the interaction strength. Although the initial inhomogeneity eventually disappears when the interaction is finite, a more significant slowing down is observed for weaker interactions in the relaxation dynamics. Remarkably, such localization behavior is obtained only when the ratio of the two types of hopping in the sawtooth lattice is close to the value at which the flat band is formed in the single-particle excitation spectrum. We also perform the level-spacing analysis via a large-size exact diagonalization using the QS$^3$ package \cite{ueda2022quantum} to examine the integrability of the model with different values of the parameters. 

\section{\label{sec:model}Model}

\begin{figure}
    \centering
    \includegraphics[width=8cm]{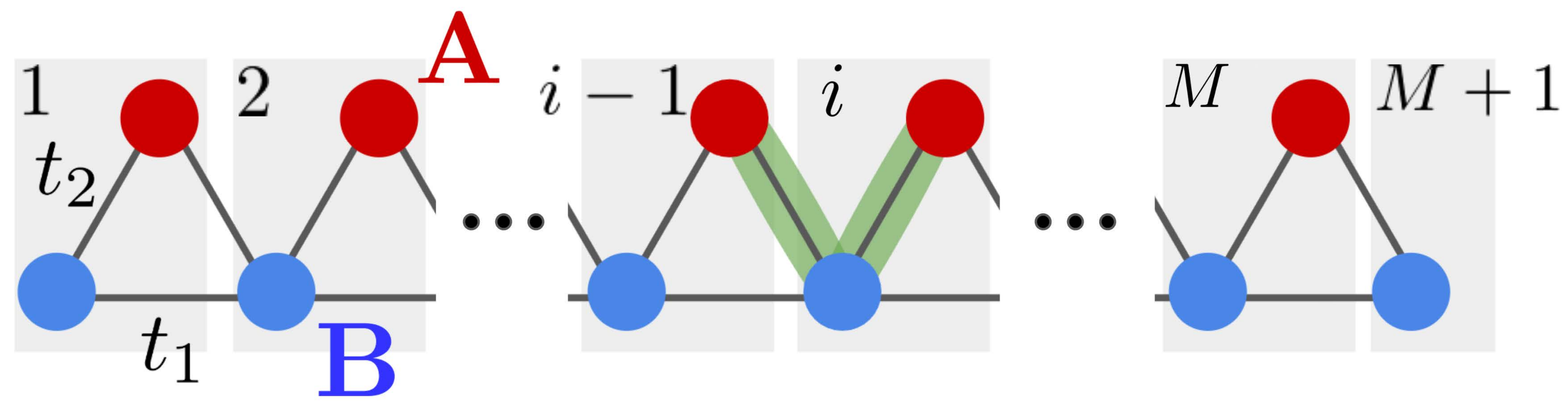}
\caption{(Color online) Schematic representation of the model considered in this study. Bosons on the lattice feel onsite repulsion with strength $U$. The hopping amplitudes between neighboring $B$ (blue) sites and those between $B$ (blue) and $A$ (red) are $t_1$ and $t_2$, respectively. Unit cells are represented by gray rectangles. The model has $L=2M+1$ sites and $M+1$ unit cells, where $A$ site is missing in the $(M+1)$-th unit cell. For $t_1/t_2 = 1/\sqrt{2}$, the model has single-particle localized eigenstate $|V_i\rangle$  represented in green.}
    \label{fig:sawtooth_fig}
\end{figure}

In this work, we consider the Bose-Hubbard model on the sawtooth lattice shown in Fig. \ref{fig:sawtooth_fig}. The Hamiltonian is expressed as
\begin{equation} \label{hamiltonian}
\begin{split}
    \hat{H} = \hat{H}_{\rm hop} + \hat{H}_{\rm int}
\end{split}
\end{equation}
with
\begin{equation} 
\begin{split}
    \hat{H}_{\rm hop} &= \sum_{i=1}^{M} [t_2(\hat{b}_{i,A}^\dagger \hat{b}_{i,B} + \hat{b}_{i,A}^\dagger \hat{b}_{i+1,B}) + t_1\hat{b}_{i,B}^\dagger \hat{b}_{i+1,B} + h.c] \\
    \hat{H}_{\rm int} &=  \frac{U}{2} [ \sum_{i=1}^{M} \hat{n}_{i,A}(\hat{n}_{i,A} - 1) + \sum_{i=1}^{M+1} \hat{n}_{i,B}(\hat{n}_{i,B} - 1)],
\end{split}
\end{equation}
where $\hat{b}_{i,\mu}$ ($\mu = A,B$) is the bosonic annihilation operator at the site $\in \mu$ in the $i$-th unit cell, we refer site $(i, \mu)$, $t_1$ and $t_2$ are hopping amplitudes between neighboring $B$ sites and those between $B$ site and $A$ site, respectively, and $\hat{n}_{i,\mu}= \hat{b}_{i,\mu}^\dagger \hat{b}_{i,\mu} $ is the site occupation operator at the $(i, \mu)$ site. The total number of sites is $L=2M+1$, and we employ open boundary condition when not explicitly stated. In this work, we consider positive hopping amplitudes, $t_{1,2} \geq 0$, which are realizable, e.g., by introducing a fast oscillation of the lattice \cite{lignier2007dynamical,eckardt2005superfluid} or by preparing a negative-temperature state \cite{braun2013negative,yamamoto2020frustrated}.

For the special ratio $t_1/t_2 = 1/\sqrt{2} \simeq 0.707$, the hopping part of the Hamiltonian $\hat{H}_{\rm hop}$ possesses localized kinetic energy eigenstates, and exhibits a flat band in the dispersion relation under periodic boundary conditions \cite{Huber2010-bw}. For example, we can construct ``V''-shaped localized single-particle eigenstate over three sites, shown in green in Fig. \ref{fig:sawtooth_fig}, with energy $E =-\sqrt{2}t_2$ \cite{Huber2010-bw}:
\begin{equation} \label{localized_state}
    |{V_i}\rangle = \hat{V}_i^\dagger \ket{0} = \frac{1}{2}(\sqrt{2}\hat{b}_{i,B}^\dagger - \hat{b}_{i-1,A}^\dagger - \hat{b}_{i,A}^\dagger) \ket{0}.
\end{equation}
Since these states $|{V_i}\rangle$ are linearly independent of each other and complete in the flat band subspace, all eigenstates of $\hat{H}_{\rm hop}$ with the same energy $E=-\sqrt{2}t_2$ are given as a superposition of these $|{V_i}\rangle$'s at different sites.
The operator $\hat{V}_i^\dagger$ and the hopping Hamiltonian $\hat{H}_{\rm hop}$ satisfy $[\hat{H}_{\rm hop}, \hat{V}_i^\dagger] = -\sqrt{2}t_2\hat{V}_i^\dagger$. Therefore, an $N$-particle localized state $\hat{V}_{i_1}^\dagger \hat{V}_{i_2}^\dagger...\hat{V}_{i_N}^\dagger \ket{0}$ is also an eigenstate of $\hat{H}_{\rm hop}$ with energy $E = -\sqrt{2}t_2N${, which is easily derived as: 
\begin{eqnarray}
\hat{H}_{\rm hop}\hat{V}_{i_1}^\dagger \hat{V}_{i_2}^\dagger...\hat{V}_{i_N}^\dagger \ket{0}\!\!\!\!&=&\!\!\!\!\hat{V}_{i_1}^\dagger (\hat{H}_{\rm hop}-\sqrt{2}t_2)\hat{V}_{i_2}^\dagger...\hat{V}_{i_N}^\dagger \ket{0}\nonumber \\
\!\!\!\!&=&\!\!\!\!\hat{V}_{i_1}^\dagger \hat{V}_{i_2}^\dagger(\hat{H}_{\rm hop}-2\sqrt{2}t_2)...\hat{V}_{i_N}^\dagger \ket{0}\nonumber \\
\!\!\!\!&=&\!\!\!\!\hat{V}_{i_1}^\dagger \hat{V}_{i_2}^\dagger...\left(\hat{H}_{\rm hop}-\sqrt{2}t_2(N-1)\right)\hat{V}_{i_N}^\dagger \ket{0} \nonumber \\
\!\!\!\!&=&\!\!\!\!-\sqrt{2}t_2N\hat{V}_{i_1}^\dagger\hat{V}_{i_2}^\dagger...\hat{V}_{i_N}^\dagger \ket{0}.
\end{eqnarray}
}

 For finite repulsion $U>0$, the commutation relation becomes
 \begin{equation} \label{H_commutator}
 [\hat{H}, \hat{V}_i^\dagger] = -\sqrt{2}t_2\hat{V}_i^\dagger + U(\sqrt{2}\hat{b}_{i,B}^\dagger \hat{n}_{i,B} - \hat{b}_{i,A}^\dagger \hat{n}_{i,A} - \hat{b}_{i-1,A}^\dagger \hat{n}_{i-1,A}).    
 \end{equation}
 Due to the appearance of the second term in Eq. \eqref{H_commutator}, the $N$-particle localized state $\hat{V}_{i_1}^\dagger \hat{V}_{i_2}^\dagger...\hat{V}_{i_N}^\dagger \ket{0}$ is no longer exact eigenstate of $\hat{H}$ when there exist ``V''s overlapping each other. Note that a pair of neighboring ``V'' states, $\ket{V_i}$ and $\ket{V_{i+1}}$, overlap at the $(i,A)$ site, giving rise to the need to consider the effect of repulsion $U$. For $L \geq 4N + 1$, we can still construct a strictly localized eigenstate in which multiple ``V'' states are arranged without overlap. 
 {This leads to massive ground-state degeneracy in the thermodynamic limit ($L\rightarrow \infty$), due to the numerous possible configurations of non-overlapping localized states for fillings up to 1/4~\cite{Huber2010-bw}. At 1/4 filling, a charge-density-wave phase forms, with the saw-tooth lattice closely packed with non-overlapping localized states}.

\section{\label{sec:single-dy}Single-Particle Dynamics}
\begin{figure}
    \centering
    \includegraphics[width=8.5cm]{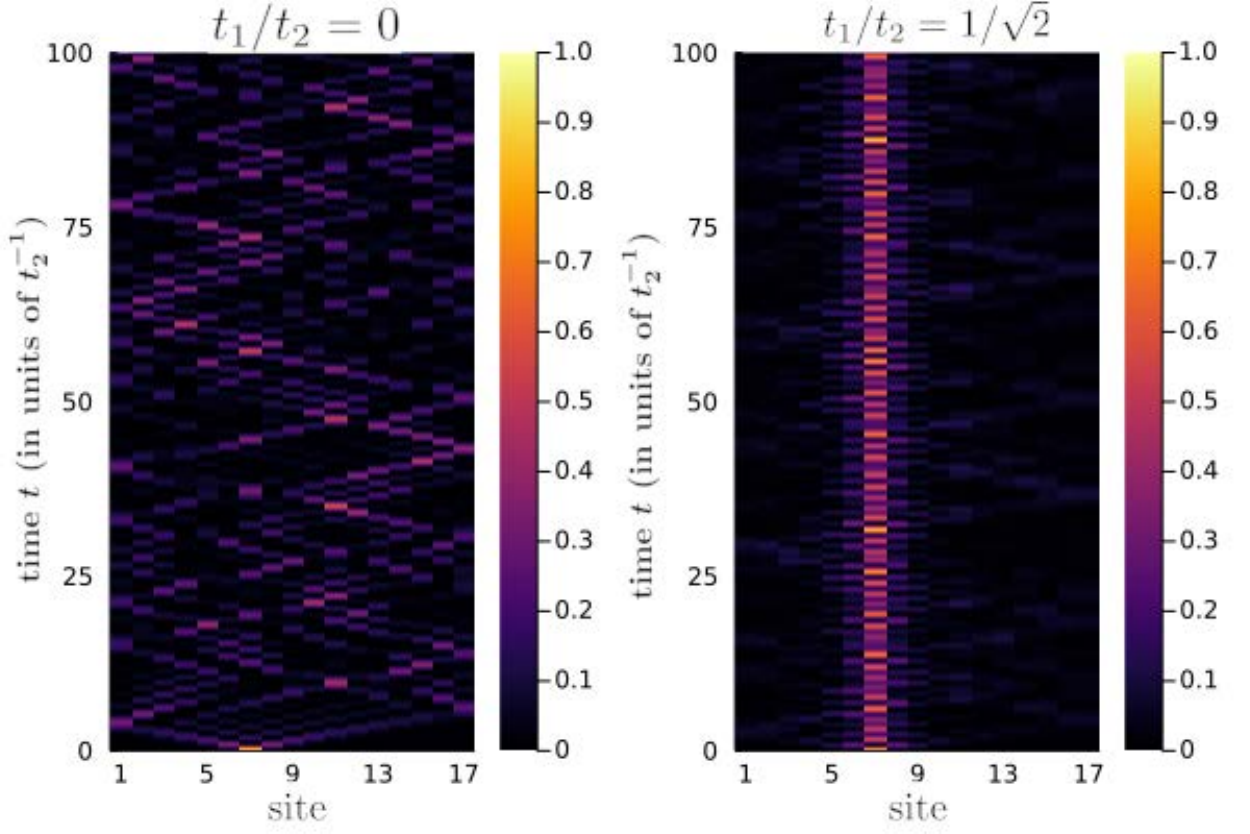}
    \caption{(Color online) The single-particle density dynamics for the initial state in which a particle is placed at the $(4, B)$ site with the system size $L = 17$. The color brightness corresponds to particle density, with brighter hues indicating higher density. The left panel corresponds to the chain limit, where $t_1/t_2 = 0$, and the right panel represents the flat band limit with $t_1/t_2 = 1/\sqrt{2}$. The site indices on the horizontal axis are allocated based on the following rule: $(1,A) \rightarrow 1$, $(1,B) \rightarrow 2$, $(2,A) \rightarrow 3$, and so forth. }
    \label{fig:dynamics_heatmap_1p_sawtooth}
\end{figure}
The existence of the localized eigenstates is expected to affect the dynamics of the system. 
Let us first consider the case of a single particle on the lattice, where the interaction $U$ does not work. In Fig. \ref{fig:dynamics_heatmap_1p_sawtooth}, we present the time evolution of particle density at each site, referred to as density dynamics, for two distinct cases: the chain limit ($t_1/t_2 = 0$) and the flat band limit ($t_1/t_2 = 1/\sqrt{2}$). The left panel, corresponding to the the chain limit, shows that the particle moves and spreads out by hoppings from site to site, as naturally expected. Conversely, the right panel, corresponding to the flat band limit, exhibits particle localization. This can be attributed to the existence of the localized eigenstates as explained below.

We elucidate the process through which a single particle becomes localized at the flat band limit $t_1/t_2 = 1/\sqrt{2}$. Arbitrary single-particle state can be expressed using a set of complex coefficients, denoted by $v_i$ and $c_n$, as follows:
\begin{equation}
    \ket{\psi} = \sum_{i=2}^{M} v_i \ket{V_i} + \sum_{n=1}^{M+2} c_n \ket{E_n},
\end{equation}
 where $\ket{E_n}$ is nonlocalized eigenstate with eigenvalue $E_n$ and assumed to be not degenerate for simplicity. The following discussion of localization can also be applied to the cases where degeneracy is present. The time evolution of a local observable, $\hat{O}$, with respect to $\ket{\psi(t)} = \exp(-iHt)\ket{\psi}$, is given by
\begin{multline}
    \braket{\hat{O}(t)} \equiv \braket{\psi (t)|\hat{O}|\psi (t)} \\
= \sum_{ij} v_i^* v_j \braket{V_i|\hat{O}|V_j} + \sum_{nm} c_n^* c_m \braket{E_n|\hat{O}|E_m} e^{-i(E_m - E_n)t} \\ + 2 \mathrm{Re}[\sum_{i,n} c_n^* v_i \braket{E_n|\hat{O}|V_i} e^{-i(E - E_n)t}]
\end{multline}
since the states $\ket{V_i}$ have the same energy $E = - \sqrt{2}t_2 \neq E_n$, for all $n$, regardless of $i$.
The infinite time average of $\braket{\hat{O}(t)}$ becomes 
\begin{multline} \label{O_inf_ev}
    \overline{\braket{\hat{O}(t)}} \equiv \lim_{\tau\rightarrow \infty} \frac{1}{\tau} \int_0^\tau \braket{\hat{O}(t)} dt 
    = \sum_i |v_i|^2 \braket{V_i|\hat{O}|V_i} \\ + \sum_{i\ne j} v_i^* v_j \braket{V_i|\hat{O}|V_j} + \sum_n |c_n|^2 \braket{E_n|\hat{O}|E_n}.
\end{multline}
Equation \eqref{O_inf_ev} shows that the coefficients of eigenstates at the initial time, denoted by $c_n$ and $v_n$, influence the infinite-time evolution. {This mechanism appears to share some similarities with the so-called Aharonov-Bohm caging~\cite{AB1,AB2,AB3}}.

\begin{figure}
    \centering
    \includegraphics[width=7.5cm]{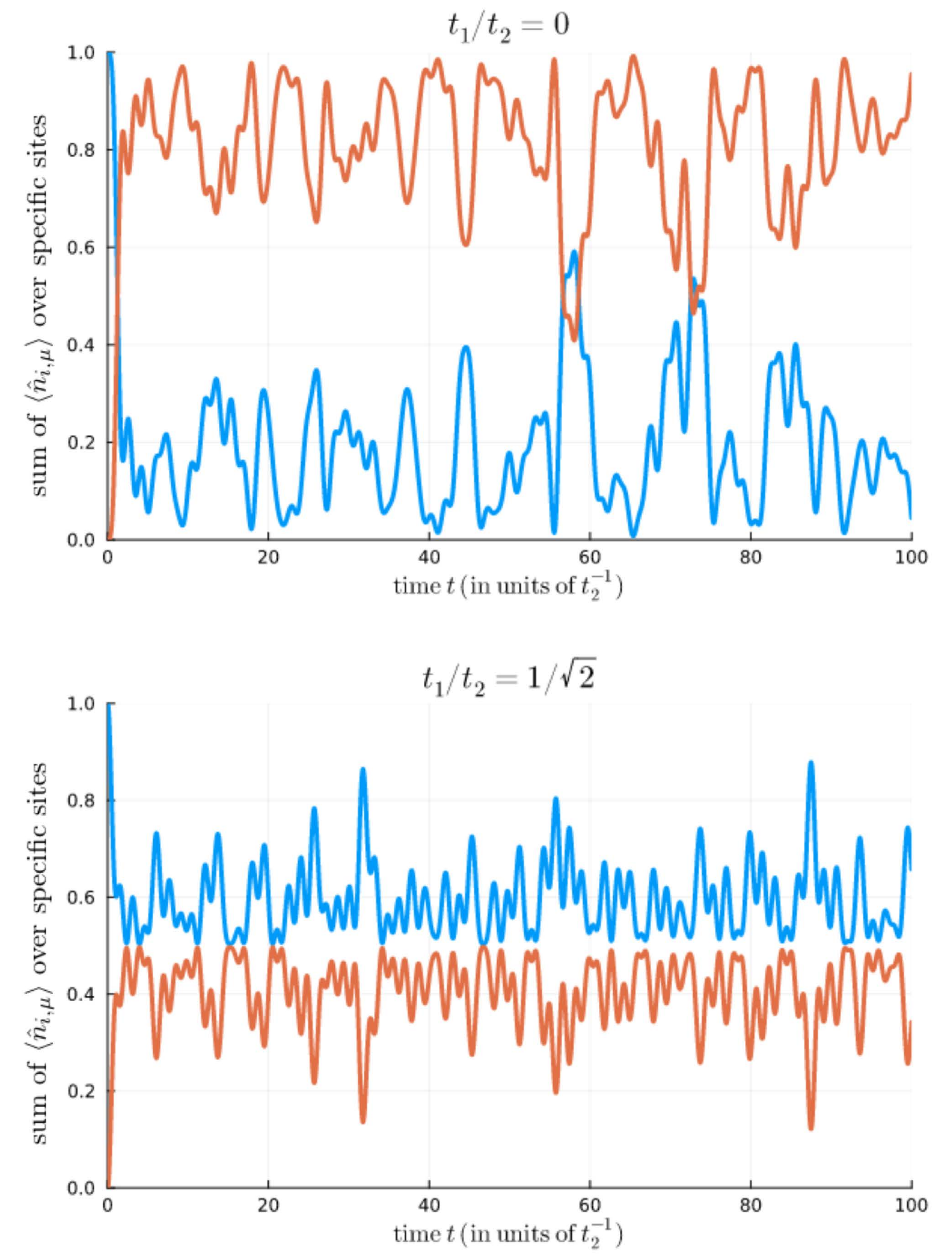}
    \caption{(Color online) The single-particle density dynamics for the initial state in which a particle is placed at the $(4, B)$ site with the system size $L = 17$. The blue lines represent the dynamics of the sum of the particle densities at sites $(3, A)$, $(4, B)$, and $(4, A)$, while the red lines represent those at the remaining sites. The top and bottom panels are for the chain limit $t_1/t_2 = 0$ and the flat band limit $t_1/t_2 = 1/\sqrt{2}$, respectively.}
    \label{fig:dynamics_1p}
\end{figure}

 Let us consider $\hat{n}_{i,\mu}$ ($\mu = A, B$) as the observable and $\ket{j,B} = \hat{b}_{j,B}^\dagger \ket{0}$ as the initial state. Since $\ket{j,B}$ has only a $\ket{V_j}$ component among the V-shaped localized eigenstates and its coefficient $v_j$ is $\braket{V_j|j, B}=1/\sqrt{2}$, the infinite-time average becomes 
\begin{multline}
    \overline{\braket{\hat{n}_{i,\mu}}}
    = \frac{1}{2} \braket{V_j|\hat{n}_{i,\mu}|V_j} + \sum_n |c_n|^2 \braket{E_n|\hat{n}_{i,\mu}|E_n}.
\end{multline}
The first term for $(i, \mu)=(j,B)$ [resp. for $(i,\mu)=(j,A)$ or $(j-1, A)$] is obtained to be $1/4$ [resp. $1/8$].  Therefore, it is concluded that, at least, $1/4 + 2 \times 1/8 = 1/2$ of the particle density remains at the three neighboring sites, $(j,B)$, $(j,A)$, $(j-1,A)$, persistently. This is confirmed in Fig. \ref{fig:dynamics_1p}, which displays the particle density at both the chain limit $t_1/t_2 = 0$ and the flat band limit $t_1/t_2 = 1/\sqrt{2}$ in the top and bottom panels, respectively. The blue lines indicate the dynamics of the sum of the particle density at sites $(j-1, A)$, $(j, B)$, and $(j, A)$, while the red lines represent the dynamics of the sum at the remaining sites. In contrast to the chain limit displayed in the top panel, the bottom panel of the flat band limit reveals that over half of the particle density is localized at the aforementioned three sites. Although we show here an example of the initial state with a particle placed at site $B$, the same conclusion can be drawn for a particle initially placed at site $A$. As shown above, the existence of a localized eigenstate can suppress the diffusion of the single particle.

\section{\label{sec:few-dy}Two-Particle Dynamics}
\begin{figure}
    \centering
    \includegraphics[width=8.5cm]{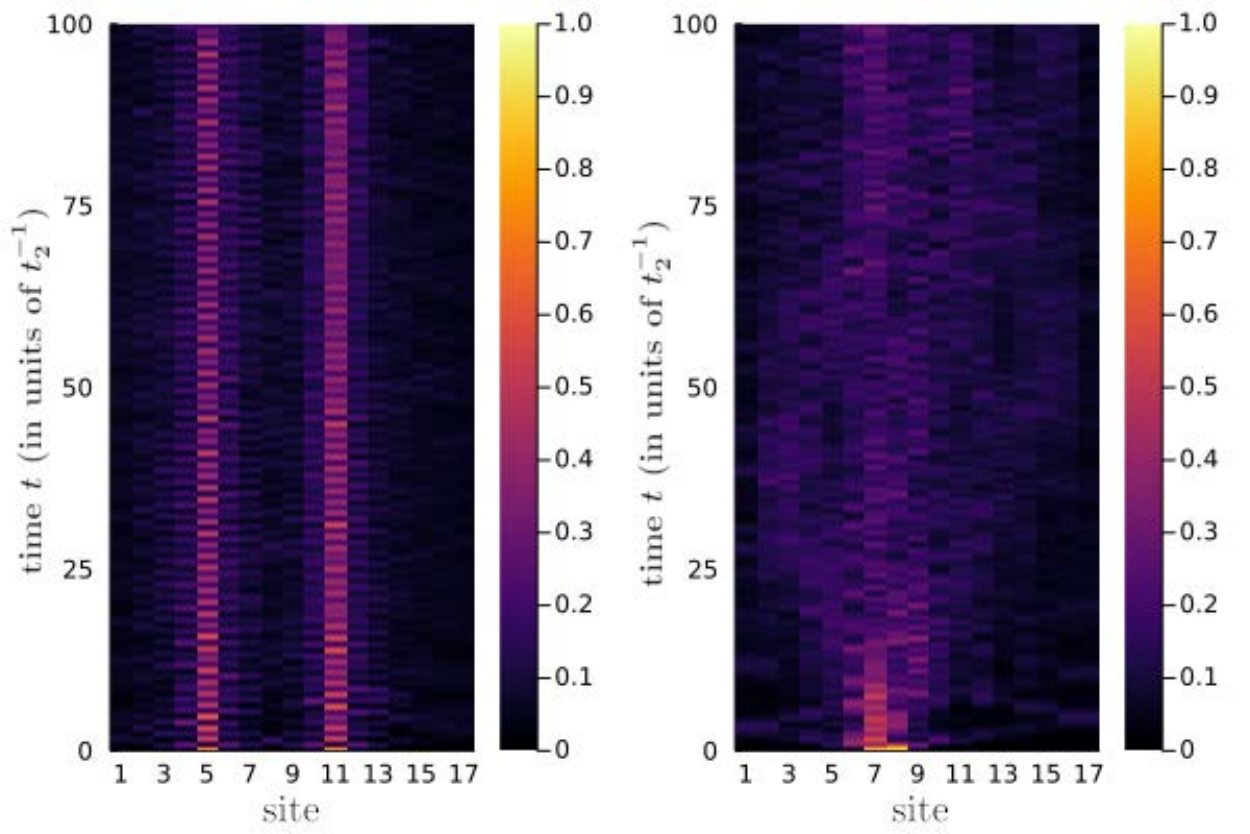}
    \caption{(Color online) Density dynamics at the flat band limit $t_1/t_2 = 1/\sqrt{2}$ for $U/t_2 = 1$, $N=2$ particles, and the system size $L=17$. The color brightness corresponds to the particle density, with brighter hues indicating higher density. The left panel is for the initial state with particles placed at sites $(3, B)$ and $(6, B)$ and the right panel is for the initial state with particles placed at sites $(4, A)$ and $(4, B)$. The site indices on the horizontal axis are allocated based on the following rule: $(1,A) \rightarrow 1$, $(1,B) \rightarrow 2$, $(2,A) \rightarrow 3$, and so forth. }
    \label{fig:dynamics_heatmap_2p_sawtooth}
\end{figure}
In the case of a many-body system, one must take into account the overlap of localized eigenstates and the effect of the interaction $U$ on the system's dynamics. Let us consider the case of two particles with finite repulsion $U > 0$ at the flat band limit $t_1/t_2 = 1/\sqrt{2}$. From Eq. \eqref{H_commutator}, for example, $\hat{V}_{3}^\dagger \hat{V}_{6}^\dagger \ket{0}$ is an exact eigenstate even in the presence of interaction, while those like $(\hat{V}_{4}^\dagger )^2 \ket{0}$ and $\hat{V}_{4}^\dagger \hat{V}_{5}^\dagger \ket{0}$ are no longer eigenstate. Therefore, the dynamics of the initial state with particles placed, e.g., at sites $(3,B)$ and $(6,B)$ is expected to be localized, while it should diffuse when the particles are initially placed, e.g., at sites $(4, A)$ and $(4, B)$. The time evolution of particle density in the former and latter cases with $U/t_2 = 1$ is shown in the left and right panels of Fig. \ref{fig:dynamics_heatmap_2p_sawtooth}, respectively. It is observed that the particle density spreads over the entire system only in the latter case as expected.

\section{\label{sec:results}Many-Body Dynamics}
As shown in the single-particle and two-particle cases, the localized eigenstates constructed from the flat band can inhibit diffusion. How does the system behave when more particles interact with each other? The dependence of the dynamics on the initial state is nontrivial, especially when the model is non-integrable. In what follows, we study the relaxation dynamics of interacting bosons from the point of view of the interplay among interaction $U$, integrability of the system and hopping ratio $t_1/t_2$.

 \subsection{Integrability of the Model}
\begin{figure}
    \centering
    \includegraphics[width=7.5cm]{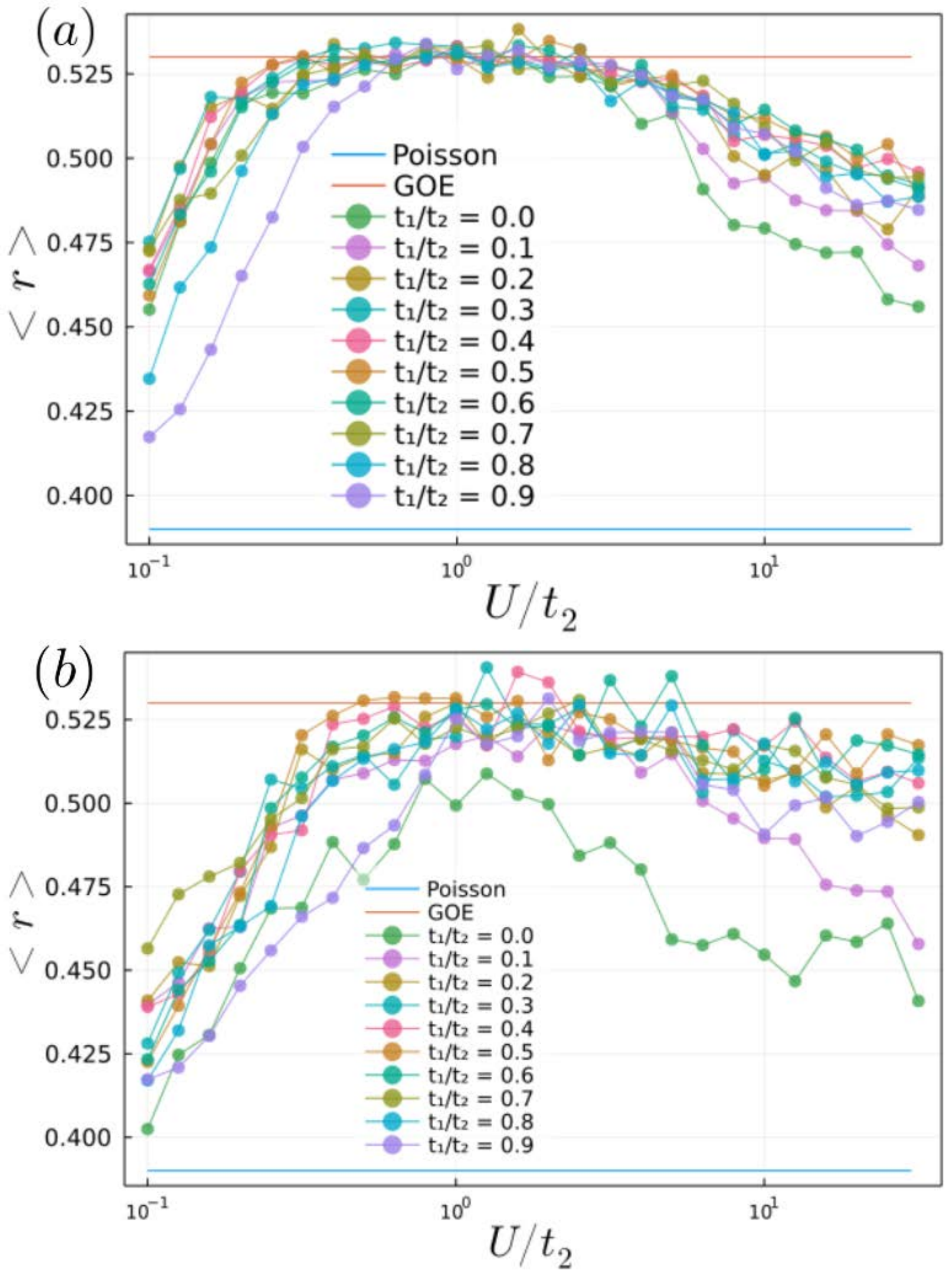}
    \caption{(Color online) Mean level spacings $\langle r \rangle$ in the {reflection-symmetric sector} as a function of the interaction $U/t_2$ for various values of $t_1/t_2$. For the Poisson distribution, $\langle r \rangle \sim 0.39$, while for the Gaussian orthogonal ensemble (GOE), $\langle r \rangle \sim 0.53$, indicating integrable and non-integrable models, respectively. (a) The system size is $L = 13$ and the numer of particles is $N = 6$. (b) The system size is $L = 17$ and the number of particles is $N = 4$. As the system size increases, these values are expected to approach the GOE values except for $U/t_2=0$ and $U/t_2 \rightarrow \infty$.  For instance, at the system size $L = 21$ and $N = 5$ particles, the value for $t_1 = 0.0$ at $U/t_2 = 1.0$ increases to 0.523.}
    \label{fig:rs}
\end{figure}
First we investigate the integrability of the model by calculating the mean level spacing $\braket{r}$, which is defined as the average (over $n$) of $r^{(n)}=\mathrm{min}(\delta^{(n)},\delta^{(n+1)})/\mathrm{max}(\delta^{(n)},\delta^{(n+1)})$, where $\delta^{(n)} = |E^{(n)} - E^{(n-1)}|$ and $E^{(n)}$ is the $n$-th eigenenergy of the model in ascending order \cite{pal2010many}. {The level spacing analysis should be conducted in a specific symmetry sector of the Hilbert space to avoid trivial degeneracy. The model under consideration exhibits U(1) symmetry, owing to particle number conservation, and Z$_2$ spatial reflection symmetry, as depicted in Fig.~\ref{fig:sawtooth_fig}. To obtain the mean level spacing $\braket{r}$ in a specific total-number sector with reflection symmetry, we calculated the energy eigenvalues $E_n$ using exact diagonalization with the QS$^3$ package \cite{ueda2022quantum}}. 

The value of $\braket{r}$ is expected to be $\braket{r} \simeq 0.53$ for non-integrable model and $\braket{r} \simeq 0.39$ for integrable model \cite{kollath2010statistical}. Figure \ref{fig:rs} plots $\braket{r}$ as a function of $U/t_2$ for several values of $t_1/t_2$, for the number of particles $N = (L - 1)/2$ and $N = (L - 1) / 4$, with system size $L$, in panels (a) and (b), respectively. At $U/t_2 \sim O(1)$, $\braket{r}$ is roughly $0.53$ indicating that the model is non-integrable, regardless of $t_1/t_2$, except for the low value of $\braket{r}$ with $t_1/t_2 = 0$ in panel (b). The value of $\braket{r}$ with $t_1/t_2 = 0$ in panel (b) is expected to increase as the size of the system increase \cite{kollath2010statistical}. For large values of $U/t_2$, it is observed that the values of $\braket{r}$ for $t_1/t_2 \leq 0.1$ are somewhat lower than those for $t_1/t_2 > 0.1$. This suggests that the presence of the $t_1$ mitigates the decrease of $\braket{r}$ caused by the effect of $U$. Given that the thermal equilibration phenomena are basically confirmed in non-integrable models \cite{deutsch2018eigenstate}, the low value of $\braket{r}$ could also hinder the relaxation of the system, regardless of whether there is a flat band or not.

\subsection{Entanglement Entropy of Eigenstates}
\begin{figure}
    \centering
    \includegraphics[width=8.cm]{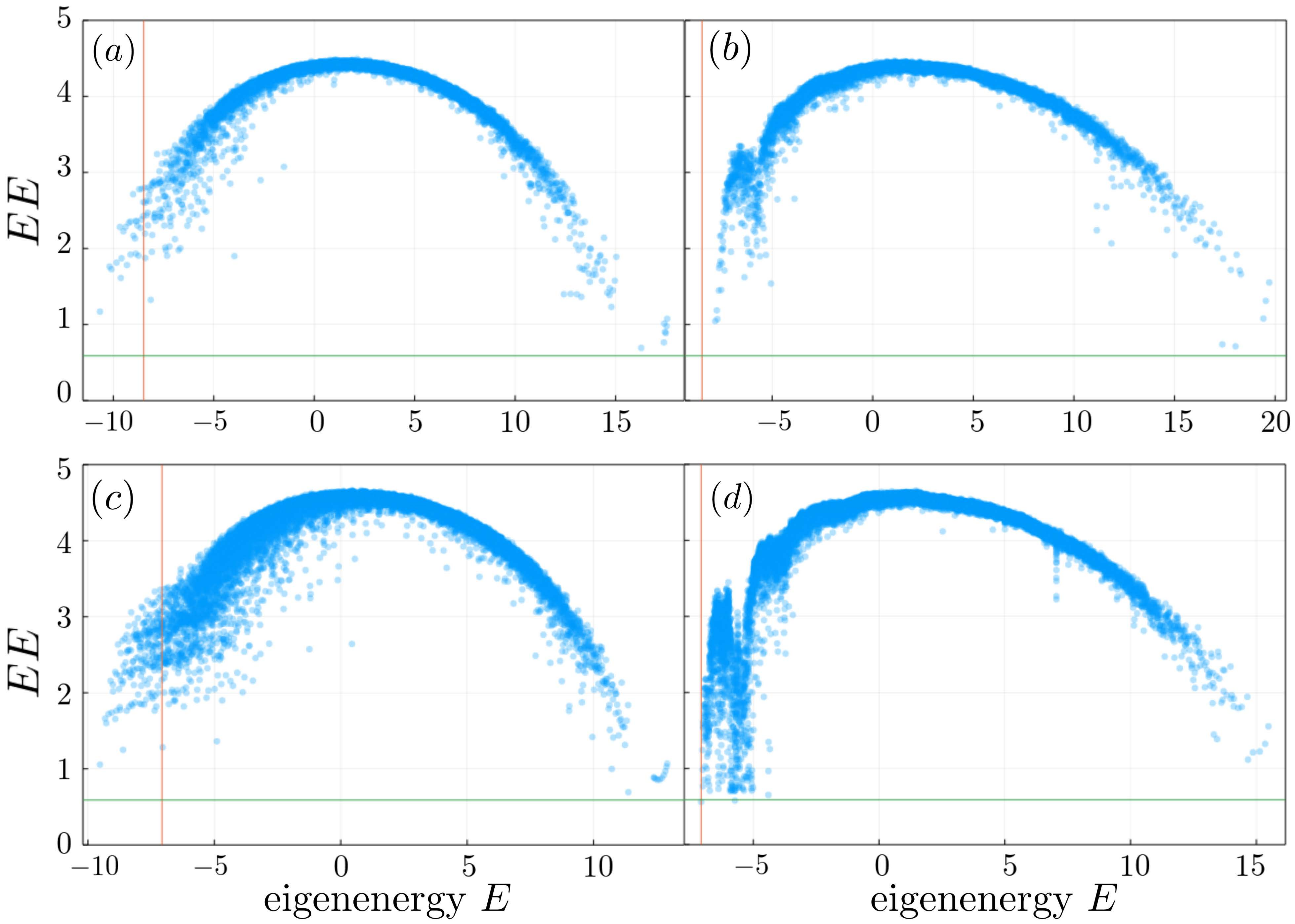}
    \caption{(Color online) Entanglement entropy (EE) between the left $(L-1)/2$ sites and right $(L+1)/2$ sites of the system for each energy eigenstates {in the reflection-symmetric sector} at $U/t_2 = 1$ (a) for the system size $L = 13$, the number of particles $N = 6$, and $t_1/t_2 = 0$ (the chain limit), (b) for $L = 13$, $N = 6$ and $t_1/t_2 = 1/\sqrt{2}$ (the flat band limit), (c) for $L = 21$, $N = 5$ and $t_1/t_2 = 0$, and (d) for $L = 21$, $N = 5$ and $t_1/t_2 = 1/\sqrt{2}$. The red line indicates the flat band energy, while the green line represents the entanglement entropy of the V-shaped localized eigenstates $\hat{V}_{i_1}^\dagger \hat{V}_{i_2}^\dagger...\hat{V}_{i_N}^\dagger \ket{0}$.}
    \label{fig:eigenstates}
\end{figure}
To study the properties of eigenstates, we investigate their entanglement entropy. For pure state $\ket{\psi}$, the entanglement entropy (EE), which is defined as $-\mathrm{Tr}_A[\rho_A \mathrm{log} \rho_A]$ with $\rho_A = \mathrm{Tr}_{\bar{A}}\ket{\psi}\bra{\psi}$, quantifies quantum correlations between the subsystem $A$ and the rest of the system $\bar{A}$. The EE of eigenstates is often studied to understand the localization properties, since a localized state is expected to have a low value of EE to reflect the paucity of distant quantum correlations.  In Fig. \ref{fig:eigenstates}, the EE between the left $(L-1)/2$ sites and right $(L+1)/2$ sites of the system, with system size $L$, for each eigenstate is displayed for the case $U/t_2 = 1$, at which the model is non-integrable. Comparing the top panels, it can be seen that for panel (a) of the chain limit, the points are scattered around a single curve, while for panel (b) of the flat band limit, it is not the case and a different structure appears at the lower energy. Similarly, the points are scattered around a single curve in panel (c) of the chain limit, while panel (d) of the flat-band limit shows many points with low EE near the red line representing the energy of the flat band $E = -\sqrt{2} t_2 N$. In particular, for $L=4N+1$, the $N$-particle localized eigenstate $\prod_{i=1}^{N}\hat{V}_{2i}^\dagger \ket{0}$, where ``V'' states are filled to maximum density without overlap, can be found as the crossing point of the red and green lines with the lowest EE in panel (d). These structures near the red line in panels (b) and (d) are considered to be remnants of the flat-band eigenstates.

\subsection{Relaxation Dynamics}
\begin{figure}
    \centering
    \includegraphics[width=8cm]{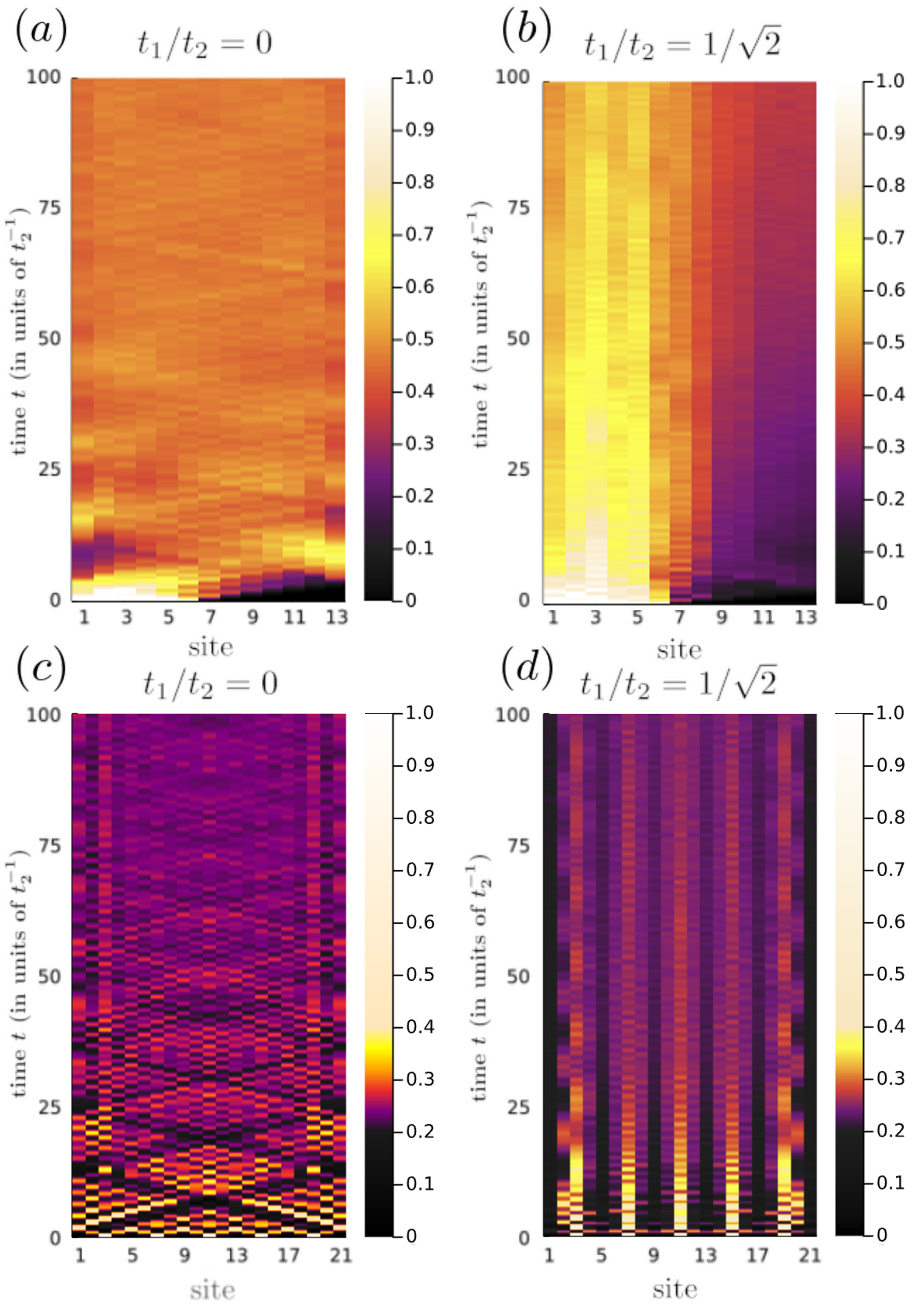}
    \caption{(Color online) Density dynamics at $U/t_2 = 1$ for the chain limit, $t_1/t_2 = 0$ [(b)(d)], and the flat band limit, $t_1/t_2 = 1/\sqrt{2}$ [(a)(c)]. The color brightness corresponds to the particle density, with brighter hues indicating higher density. [(a)(b)] The system size is $L = 13$ and the number of particles is $N = 6$. [(c)(d)] The system size is $L = 21$ and the number of particles is $N = 5$. The site indices on the horizontal axis are allocated based on the following rule: $(1,A) \rightarrow 1$, $(1,B) \rightarrow 2$, $(2,A) \rightarrow 3$, and so forth. }
    \label{fig:dynamics_heatmap}
\end{figure}
As we explained in Secs. \ref{sec:single-dy} and \ref{sec:few-dy} for the case of few particles, the existence of a flat band can affect the dynamics in a way that hinders diffusion. The remnants of the flat-band eigenstates in the non-interacting limit are expected also to influence the many-body dynamics.

Return probability $|\braket{\psi(0)|\psi(t)}|^2$ is often calculated to study the localization and preservation of initial state information in dynamics \cite{kuno2020flat,kuno2021multiple,khare2020localized}. However, in the case of softcore bosons, where the number of possible states is large, the conservation of the return probability is not expected to be a good indicator. Therefore, to examine the localization of the dynamics, we focus on the particle density dynamics also in many-body case. The density dynamics at $U/t_2 = 1$ are depicted in Fig. \ref{fig:dynamics_heatmap} for the chain limit, $t_1/t_2 = 0$, and the flat band limit, $t_1/t_2 = 1/\sqrt{2}$, on the left and right panels, respectively. In the upper panels, we consider the time evolution of the initial state with the system size $L$ and the number of particles $N = (L - 1) / 2$, where the particles are laid out in the left half of the system $\ket{\bullet \! \bullet \! \cdots \! \circ \! \circ}$. In the lower panels, with $N = (L - 1) / 4$, we consider the time evolution of the initial state where each particle is equally spaced every 4 sites $\ket{\circ \! \circ \! \bullet \! \circ \! \circ \! \circ \! \bullet \! \circ \! \cdots }$. We can see that the particle diffusion is significantly slowed down in the case of the flat band limit, compared to the chain limit, in both the top and bottom panels, even though non-integrability is expected for $U/t_2 = 1$. Having said that, as seen in Fig. \ref{fig:dynamics_heatmap}(b), the bias in the density distribution gradually diminishes when the initial state is $\ket{\bullet \! \bullet \! \cdots \! \circ \! \circ}$. In contrast, Fig. \ref{fig:dynamics_heatmap}(d) indicates that the bias seems to reduce only up to a certain level for the initial state $\ket{\circ \! \circ \! \bullet \! \circ \! \circ \! \circ \! \bullet \! \circ \! \cdots }$. 

\begin{figure}
    \centering
    \includegraphics[width=8cm]{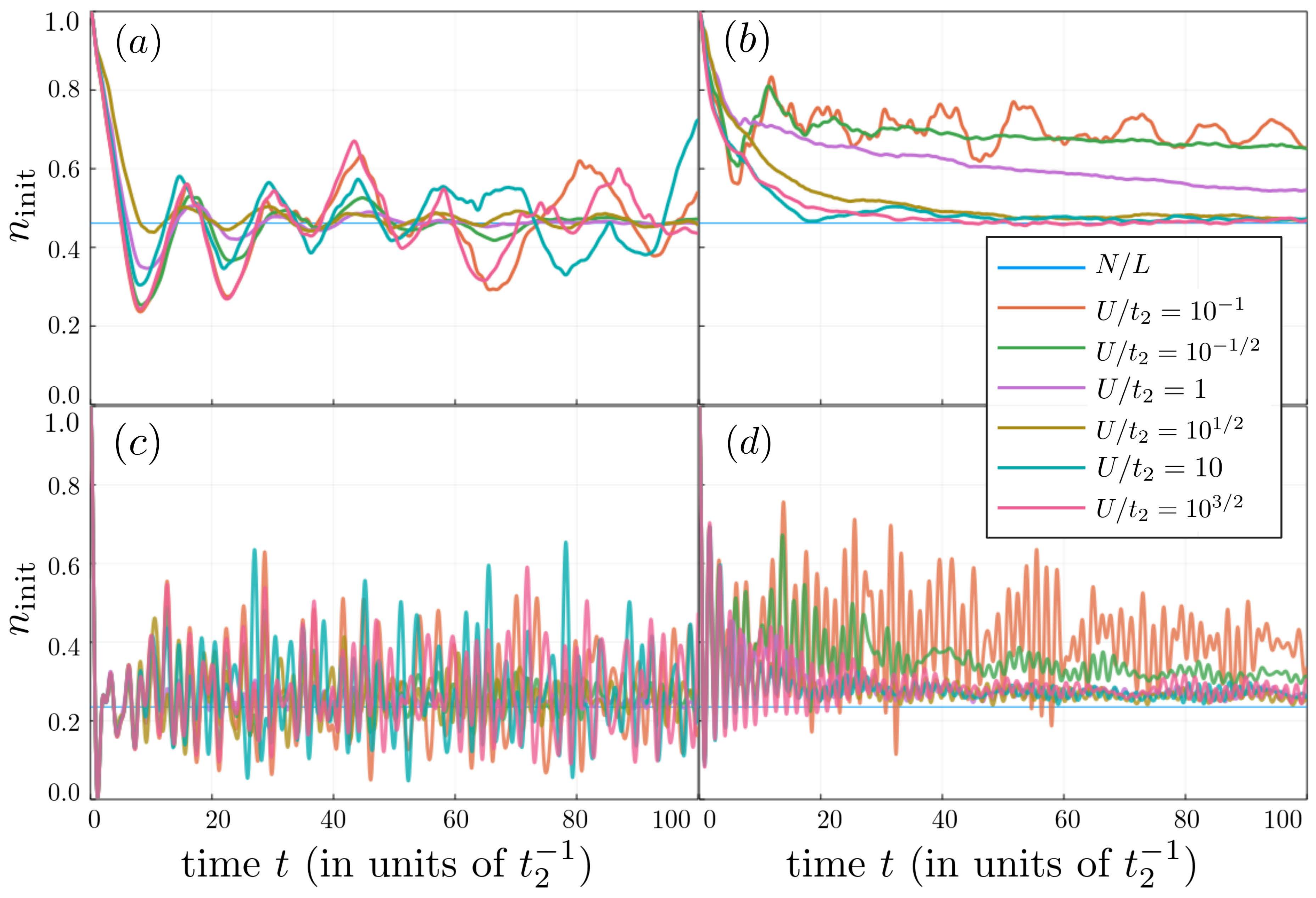}
    \caption{(Color online) Dynamics of $n_{\rm init}$, representing the average particle density at sites where particles were initially located, for various values of $U/t_2$ at the chain limit, $t_1/t_2 = 0$ [(a)(c)], and at the flat band limit, $t_1/t_2 = 1 / \sqrt{2}$ [(b)(d)]. Panels (a) and (b) are for the system size $L=13$ and $N=6$ particles, and its initial state is $\ket{\bullet \! \bullet \! \cdots \! \circ \! \circ}$. Panels (c) and (d) are for the system size $L = 17$ and $N = 4$ particles, and its initial state is $\ket{\circ \! \circ \! \bullet \! \circ \! \circ \! \circ \! \bullet \! \circ \! \cdots }$}
    \label{fig:dynamics_t_fixed}
\end{figure}
For a more direct visualization of the aforementioned observations, we plot the time evolution of $n_{\rm init}$, the average particle density at sites where particles were initially located, at various values of $U/t_2$ for the chain limit, $t_1/t_2 = 0$, and the flat band limit, $t_1/t_2 = 1/\sqrt{2}$ in Fig. \ref{fig:dynamics_t_fixed}. The initial state is $\ket{\bullet \! \bullet \! \cdots \! \circ \! \circ}$ with $N=(L-1)/2$ in Figs. \ref{fig:dynamics_t_fixed}(a) and \ref{fig:dynamics_t_fixed}(b) and $\ket{\circ \! \circ \! \bullet \! \circ \! \circ \! \circ \! \bullet \! \circ \! \cdots }$ with $N=(L-1)/2$ in Figs. \ref{fig:dynamics_t_fixed}(c) and \ref{fig:dynamics_t_fixed}(d), respectively. Interestingly, it is seen that the relaxation is considerably slow for small values of $U/t_2$ ($\leq 1$) in the flat band limit, and especially in panel (d), the bias in the distribution is not eliminated even when the value of $U/t_2$ is increased. This behavior is likely attributable to the existence of the localized eigenstate $\prod_{i=1}^{N}\hat{V}_{2i}^\dagger \ket{0}$.

\begin{figure}
    \centering
    \includegraphics[width=7.4cm]{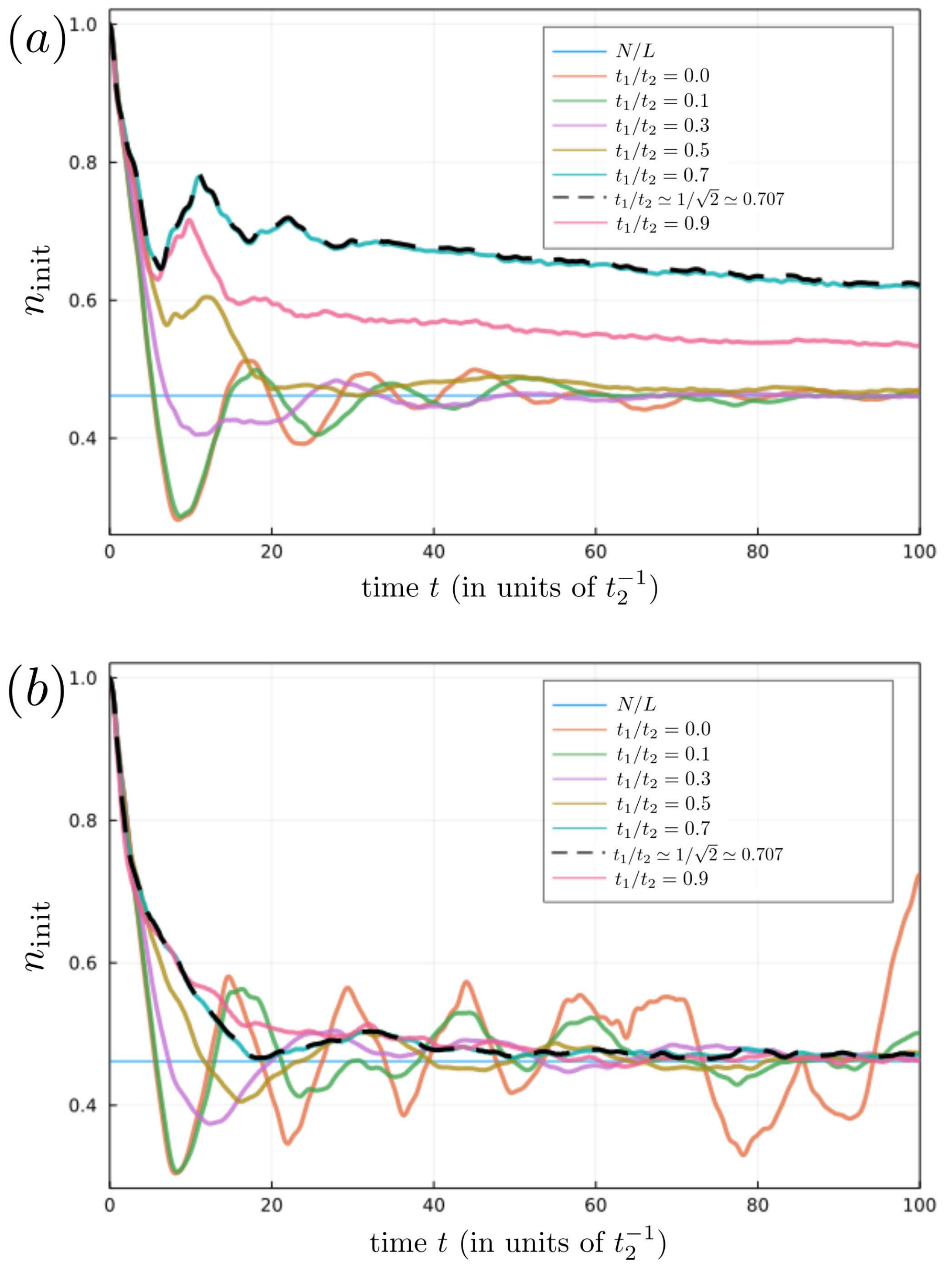}
    \caption{(Color online) Dynamics of $n_{\rm init}$, representing the average particle density at sites where particles were initially located, for various values of $t_1/t_2$ with $U/t_2 = 0.5$ (a) and $U/t_2 = 10$ (b). The system size is $L = 13$, the number of particles is $N=6$ and its initial state is $\ket{\bullet \! \bullet \! \cdots \! \circ \! \circ}$. }
    \label{fig:dynamics_u_fixed}
\end{figure}
In Fig. \ref{fig:dynamics_u_fixed}, we plot the time evolution of $n_{\rm init}$ for various values of $t_1/t_2$ with fixed $U/t_2$ in order to show the importance of the flat-band nature for the slowing down of relaxation. In Fig. \ref{fig:dynamics_u_fixed}(a), one can notice that the significant slowdown in the relaxation process is observed only for $t_1/t_2 \sim 1/\sqrt{2}$. Conversely, Fig. \ref{fig:dynamics_u_fixed}(b) indicates that the significance of the flat-band effect diminishes for large values of $U/t_2$. Note that the peculiar behavior of the relaxation dynamics for $t_1/t_2 \leq 0.1$ in Fig. \ref{fig:dynamics_u_fixed}(b) is likely a consequence of the relatively small system size, as indicated by the small values of $\braket{r}$ presented in Fig. \ref{fig:rs}. These results indicate that the presence of flat band at $t_1/t_2 = 1/\sqrt{2}$ limits the diffusion of particle density more significantly for smaller $U/t_2$.

\section{\label{sec:conclusion}Conclusion}
In this paper, we have studied the relaxation dynamics of the Bose-Hubbard model on the sawtooth lattice, where a flat band appears in the single-particle excitation spectrum when the ratio of the two types of hopping is tuned to a special value. We first showed that in the one-particle case, localized eigenstates in the flat-band limit induce particle localization during time evolution. We then considered the two-particle case and confirmed that particle interactions can hinder this localization. 

For many-body systems with interactions, we first performed numerical calculations of the eigenstates with exact diagonalization using the QS$^3$ package \cite{ueda2022quantum}. The eigenstates of the system and their level spacings provided insight into the degree of integrability of the system. Additionally, we computed the entanglement entropy, which signifies the presence of localized eigenstates in the flat-band case. With these insights into the eigenstates in hand, we proceeded to study the time evolution of the system from two different initial conditions, employing the TEBD method \cite{paeckel2019time}. {For an initial state with particles arranged on one half of the system, the inhomogeneity eventually dissipated. However, this process took significantly longer, particularly in cases of weaker interactions when the hopping ratio $t_1/t_2$ approaches the flat band limit. This prolonged duration can be attributed to the presence of numerous localized eigenstates at the non-interacting limit.} For the initial state where the density is one particle per four sites and each particle is equally spaced, the inhomogeneity of the state did not completely disappear even after a long time and even with strong interaction. {This persistence can be attributed to the presence of a localized eigenstate that remains unaffected by the interactions.} 

{The system studied here solely considers on-site repulsion as an interaction, with our main focus directed towards the particle number density in the dynamics. Consequently, experimental verification of these findings and exploration of cases with larger system sizes not accessible through numerical calculations could be achieved using bosonic ultracold atoms in an optical sawtooth lattice \cite{zhang2015one}. The two initial states considered here could be prepared by applying techniques demonstrated in previous studies~\cite{Trotzky2012,choi2016exploring}. Furthermore, our bosonic flat-band model could be extended to include a broader range of lattice geometries, such as kagome \cite{Jo2012} and Lieb \cite{taie2015} optical lattices. With experimental feasibility in mind, the present study establishes a robust foundation for quantum simulation studies on the dynamics of systems exhibiting atypical eigenstates. This particularly encourages further comprehensive examinations of the interplay effects among the flat band resulting from lattice geometry, the softcore nature of bosons, and the prepared initial states, in relation to the eigenstate thermalization hypothesis.}

\begin{acknowledgments}
We would like to thank H. Ueda for his invaluable assistance regarding the calculations using the QS$^3$ package.
This work was supported by JSPS KAKENHI Grant Nos. 21H05185 (DY), {23K22442} (NF, DY), {23K25830} (DY), {24K06890 (DY)}, and JST PRESTO Grant No. JPMJPR2118, Japan (DY).
\end{acknowledgments}

\end{document}